# Two characteristic constants of the supercooled liquid transitions of amorphous substances

Wenlong Jiang[1)†]
1) Independent Researcher, Chizhou, Anhui Province, People's Republic of China
Dated: December 14, 2024

**ABSTRACT**

Supercooled liquid state is a particularly interesting state in that it exhibits several unusual physical properties. To illustrate, the liquid displays a single peak relaxation frequency at high temperatures, which splits into $\alpha$ relaxation and $\beta$ relaxation in the moderately supercooled regime, with relaxation a disappearing at the glass transition temperature. The mechanism underlying these unusual physical properties of liquids has always been one of the important research topics in condensed matter. Here, a new mechanism is proposed. A distinctive physical state is built, and its most salient feature is that its independent variables are difficult or impossible to measure. Theoretical calculations indicate that there exist two sets of measurable variables in this physical state that cannot be measure exactly simultaneously. Moreover, it is easy to reach an erroneous conclusion, namely that "a system in this physical state is in a superposition of some real states, until it is measured". Further theoretical calculations demonstrate that there are two new transitions and that $e^3$ and $2e^3$ are characteristic values of these two transitions, respectively, where e is Euler's number. Considerable experimental data shows that the characteristic value of glass transition appears to be concentrated near $2e^3$ and the characteristic value of another transition (for example, the splitting of relaxation peak) appears to be concentrated near $e^3$.

**† Corresponding Author.**
E-mail: guihuashu@126.com, ORCID ID:0000-0003-0310-1205.

# 1 INTRODUCTION

Glassy substances are widely used in Industry, home and daily life, such as plastics, bottles, asphalt, rubber and optical fibres. Understanding and controlling it is important for many materials, such as Glass fibre sizing[1], glass-ceramic photocatalysts Architectural Glass[2], Metallic glass[3], Glass Fibers Reinforced Concrete[4], and many others. A glass is usually made by cooling a liquid fast enough to prevent crystallisation[5,6]. Although this method of making glass has been known for thousands of years, the process by which the liquid assumes its glassy state as it cools has not yet been fully understood. The phenomena that occur during the rapid cooling of liquids, and the mechanisms of these phenomena are important research areas in condensed matter. A great deal of experimental and theoretical research has been conducted on this process, and many important conclusions have been drawn. Here we are more concerned with those studies that obtain a constant. Because the constant is determined by the nature of the transition from liquid to glass, it may be directly correlated with or reflective of the nature of the transition. Some of constants pertinent to the phenomenon of rapid cooling of liquids are listed below.

It is generally accepted that the viscosity of a liquid reaches a critical value of $10^{13}$ poise at the glass transition temperature ($T_g$), and that this value is independent of the amorphous component[7,8,9,10].

In 1955, M. L. Williams, R. F. Landel and J. D. Ferry proposed the now well-known WLF equation,

$$\log a_T = -\frac{\lambda_1(T-T_g)}{\lambda_2+T-T_g}, \qquad (1)$$

where $\lambda_1$ and $\lambda_2$ are constants, $a_T$ is a ratio and T is the temperature. The equation is applicable to temperatures within the range of $T_g$ and $T_g + 100°C$, where °C is degree Celsius. $\lambda_1$ appears to have universal values of 17.44. Therefore, the WLF fractional free volume at the glass transition is approximately 0.025[11].

By studying the change in RFV (reduced free volume) during yielding for dozens of different amorphous alloys, Wang Weihua et al. observed that the RFV displays a pronounced surge when the RFV reaches a critical value of approximately 2.4% [12]. This numerical result is independent of the chemical composition and mechanical properties of the amorphous alloy, as well as the activation energy required for atomic transitions[12]. In other words, the critical free volume at the onset of yielding for these amorphous alloys is typically approximately 2.4%.

Robert Simha and R. F. Boyer observed that the difference between the volume expansion coefficients of the liquid ($a_L$) and the volume expansion coefficients of glassy states ($a_g$) is inversely proportional to the glass transition temperature for a polymer. This relationship can be expressed as $(a_L - a_g)T_g \approx 0.113$[13,14,15,16], where $\approx$ approximately "equal to". Therefore, it can be concluded that $(a_L - a_g)T_g$ of polymers is approximately 0.113. Bo Shi et al. found a similar relationship. They found that the product of the average linear thermal expansion coefficient of metallic glasses and the glass transition temperature $T_g$ is essentially independent of $T_g$ and equal to $0.75 \times 10^{-2}$[17]. Peter Lunkenheimer et al. found that $a_l/a_g$ is on the order of 3 for a large amount of glass-forming systems, where $a_l$ is the thermal expansion coefficient in liquids and $a_g$ is the thermal expansion coefficient in glasses[18].

K. L. Ngai and S. Capaccioli observed that the $\frac{E_\beta}{RT_g}$ of many nonmetallic materials is around 24, where $E_\beta$ is the activation energy of the $\beta$ relaxation and R is the gas constant[19]. The $\frac{E_\beta}{RT_g}$ of many metallic glasses (MGs) has been determined by various researchers. Lina Hu, and Yuanzheng Yue determined that $\frac{E_\beta}{RT}$ of La$_{55}$Al$_{25}$Ni$_{20}$ is equal to 26.8[20]. Then, by investigating the $E_\beta$ of more than ten Mgs, they observed that the average value of $\frac{E_\beta}{RT_g}$ for these MGs is 26.1[20]. By summarising the values of $\frac{E_\beta}{RT_g}$ of a large number of MGs, Wang Weihua et al. obtained that $\frac{E_\beta}{RT_g} = 26 \pm 2$[21]. In conclusion, $\frac{E_\beta}{RT_g}$ of glass former is around 24.

The research of H. B. Yu et al. indicates that the activation of the shear transformation zone in metallic glasses is directly correlated with the $\beta$ relaxation and that the activation energy of the $\beta$ relaxation is observed to be nearly equivalent to the potential energy barriers of the shear transformation zone[21,22,23]. According to this research of H. B. Yu et al., in multielement metal glasses, the activation energy for diffusion motion of the smallest constituent atom is approximately equal to the $\beta$ relaxation activation energy[24].

At a temperature of approximately $1.2T_g$, the slow $\beta$-relaxation separates from the $\alpha$-process [25,26]. The inverse relationship between translational motion and viscosity is no longer valid when the temperature is below approximately $1.2T_g$[27]. N.A. Davydova et al. found that the temperature at which the nucleation process of m-cresol begins in the supercooled liquid is in the range of $1.29 - 1.18T_g$ [28]. The researches of these studies indicate that a change in the properties of the liquid occurs at approximately $1.2T_g$. This conclusion is also supported by the research on polymer conductivity conducted by S. H. Chung et al[29].

Interestingly, on closer examination of these constants, we can find that $\frac{1.2}{24 \times 0.025} = 2$. The simple number 2 suggest an intrinsic correlation among 0.025, 24 and $1.2T_g$. 0.025 is a constant associated with glass transition. 24 is a constant associated with $\beta$ relaxation. 1.2 is a constant associated with the temperature at which the slow $\beta$-relaxation separates from the $\alpha$-process. Therefore, the simple number 2 suggest an intrinsic correlation among glass transition, slow $\beta$-relaxation and the separation of the slow $\beta$-relaxation from the $\alpha$-process. It is unfortunately that no research has yet been conducted on this intrinsic connection. My previous research indicates that 0.025 should be $\frac{1}{2e^3}$[30]. In this paper, the author will continue the previous research and try to provide an explanation.

## 2 BASIC CONCEPTS AND METHODS

Let $F$ be $\psi(s_1, s_2, s_3, \cdots, s_N)$, where $F$ represents a physical quantity, $s_1, s_2, s_3, \cdots, s_N$ denotes N mutually independent variables, and $\psi$ is a function. Let $t_1, t_2, t_3, \cdots, t_N$ also denote N mutually independent variables, and assume that there is no relationship between $s_1, s_2, s_3, \cdots, s_N$ and $t_1, t_2, t_3, \cdots, t_N$. Let $\vec{s} = (s_1, s_2, s_3, \cdots, s_N)$ and $\vec{t} = (t_1, t_2, t_3, \cdots, t_N)$. Therefore, there is no relationship between $\vec{s}$ and $\vec{t}$. Next, we will explore how to establish a functional relationship between $\vec{s}$ and $\vec{t}$.

Because there is no relationship between $s_i$ ($i$ is a natural number and $i \leq N$) and $t_1, t_2, t_3, \cdots, t_N$, $s_i = a_i$ ($a_i$ is a real number) is equivalent to $t_1 = c_1$, or $t_2 = c_2$, $\cdots$, or $t_N = c_N$, where $c_1, c_2, \cdots, c_N$ are any real numbers. Therefore, the equation $s_i = a_i$ is equivalent to $r_i = c_i$, where $r_i$ can take on any one of $t_1, t_2, \cdots, t_N$ and $c_i$ can take on any real number. Therefore, if there is no relationship between $\vec{s}$ and $\vec{t}$, we have the equation

$(s_1 = a_1, s_2 = a_2, \cdots, s_N = a_N) \Leftrightarrow (r_1 = c_1, r_2 = c_2, \cdots, r_N = c_N)$,

where the symbol $\Leftrightarrow$ represents "equivalent to".

Since there is no relationship between $s_1, s_2, s_3, \cdots, s_N$ and $t_1, t_2, t_3, \cdots, t_N$, the values of $r_1, r_2, \cdots, r_N$ and the values of $c_1, c_2, \cdots, c_N$ are independently of each other.

Let $\vec{r}$ be $(r_1, r_2, \cdots, r_N)$. Because $r_i$ can take on any one of $t_1, t_2, t_3, \cdots, t_N$, there are $N^N$ possible $(r_1, r_2, \cdots, r_N)$ values. Let $TA$ be a space, and let the axes of $TA$ be the $t_1$-axis, $t_2$-axis, $\cdots$, and $t_N$-axis. It is obvious that a point in space $TA$ can only be of the form $(t_1 = c_1, t_2 = c_2, t_3 = c_3, \cdots, t_N = c_N)$, or $(t_2 = c_2, t_1 = c_1, t_3 = c_3, \cdots, t_N = c_N)$, or $(t_2 = c_2, t_3 = c_3, t_1 = c_1, \cdots, t_N = c_N)$, or $\cdots$. In this paper, if $(r_1 = c_1, r_2 = c_2, \cdots, r_N = c_N)$ corresponds to a point in space $TA$, we say that this $(r_1 = c_1, r_2 = c_2, \cdots, r_N = c_N)$ is valid. Therefore, the condition under which $\$(r_1 = c_1, r_2 = c_2, \cdots, r_N = c_N)$ is valid is that $r_1 \neq r_2 \neq r_3 \neq \cdots \neq r_N$. According to the above analysis, $r_i$ can take on any one of $t_1, t_2, \cdots, t_N$. Taken together, only $N!$ values are valid for $(r_1 = c_1, r_2 = c_2, \cdots, r_N = c_N)$, where $N!$ is the factorial. According to the above analysis, there are $N^N$ possible $(r_1, r_2, \cdots, r_N)$ values. Therefore, the probability that $(r_1 = c_1, r_2 = c_2, \cdots, r_N = c_N)$ is valid is $\frac{N^N}{N!}$. Let $\sigma$ represent the probability that $(r_1 = c_1, r_2 = c_2, \cdots, r_N = c_N)$ is valid. Then, we have that

$$\sigma = \frac{N!}{N^N}, \tag{2}$$

Let $S$ be a space, and let the axes of $S$ be the $s_1$-axis, $s_2$-axis, ..., and $s_N$-axis.

Let $\omega$ be a sequence of $M$ points in space $S$. Let $\mu$ be a sequence. Let $\omega \Leftrightarrow \mu$. $\omega \Leftrightarrow \mu$ means that the $i$th term of sequence $\omega$ is equivalent to the $i$th term of sequence $\mu$, where $i$ represents any natural number less than $M$. Let $RA$ be a space, and let the axes of $RA$ be the $r_1$-axis, $r_2$-axis, $\cdots$, and $r_N$-axis. Then according to Equation (2), Sequence $\mu$ should consist of $M$ points in space $RA$. According to Equation (2), when $M$ is sufficiently large, the number of elements that are valid in Sequence $\mu$ should be $M\frac{N!}{N^N}$. Therefore, a necessary condition for $\mu$ to be equivalent to $\omega$ is that Sequence $\mu$ consists of $M$ points in space $RA$ and the number of elements that are valid in Sequence $\mu$ is $M\frac{N!}{N^N}$ when $M$ is sufficiently large.

Because there is no relationship between $s_1, s_2, s_3, \cdots, s_N$ and $t_1, t_2, t_3, \cdots, t_N$, there is no universal function between $s_1, s_2, s_3, \cdots, s_N$ and $t_1, t_2, t_3, \cdots, t_N$. However, in some cases, a single sequence is equivalent to several sequences at the same time such that in these cases there exists a function between $s_1, s_2, s_3, \cdots, s_N$ and $t_1, t_2, t_3, \cdots, t_N$. Therefore, we can build a special case, in which $t_1, t_2, t_3, \cdots, t_N$ is a function of $s_1, s_2, s_3, \cdots, s_N$, by following the steps below.

**Step 1** *Let Sequence $A$ be the sequence consisting of all possible values of $s_1, s_2, s_3, \cdots, s_N$. Hence, Sequence $A$ is composed of $M$ different points in space $S$, and $M$ should be a number approaching infinity.*

**Step 2** *Sequence $A$ can be divided into $\frac{N^N}{N!}$ sequences, where sequence $A_z$ is composed of elements in sequence $A$ located at positions $zM\frac{N!}{N^N}+1$, $zM\frac{N!}{N^N}+2$, $zM\frac{N!}{N^N}+3$, $\cdots$, $(z+1)M\frac{N!}{N^N}$, where $z$ is a natural number and $z \leq \left(\frac{N^N}{N!}-1\right)$.*

**Step 3** *Let sequence $B_z$ is composed of $M$ different points in space $RA$; the number of elements that are valid in Sequence $B_z$ is $M\frac{N!}{N^N}$ and the positions of these valid elements in Sequence $B_z$ are $zM\frac{N!}{N^N}+1$, $zM\frac{N!}{N^N}+2$, $zM\frac{N!}{N^N}+3$, $\cdots$, $(z+1)M\frac{N!}{N^N}$. It is evident that sequence $A$ and sequence $B_z$ satisfy Equation (2), so $A \Leftrightarrow B_z$. Because $A \Leftrightarrow B_z$, there exists a function $H_z$ such that $\vec{s}=H_z(\vec{r})$ and $\vec{r}=H_z^{-1}(\vec{s})$, where $\vec{r} \in B_z$, $\vec{s} \in A$ and the symbol $\in$ represents "belong to". $H_z^{-1}$ is the invertible function of $H_z$. Sequence $C_z$ is composed of valid elements in sequence $B_z$, in the order in which they appeared in the sequence $B_z$. It is evident that $\vec{s}=H_z(\vec{t})$ and $\vec{t}=H_z^{-1}(\vec{s})$, where $\vec{t} \in C_z$ and $\vec{s} \in A_z$.*

**Step 4** *An element can be observed only when it is valid. Conversely, an element cannot be observed when it is not valid.*

**Step 5** *Sequence $C_1$, sequence $C_2$, $\cdots$, sequence $C_z$, $\cdots$, sequence $C_{\frac{N^N}{N!}}$ Satisfy*

$$C_1 = C_2 = \cdots = C_z = \cdots = C_{\frac{N^N}{N!}}.$$

**Step 6** *All possible values of $\vec{s}$ have an equal probability of occurring. That is, $s_1, s_2, s_3, \cdots, s_N$ can assume any of the possible values freely.*

**Step 8** *$\vec{t}$ is continuous.*

Sequence $C$ is composed of sequence $C_1$, sequence $C_2$, sequence $C_3$, $\cdots$, sequence $C_{\frac{N^N}{N!}}$ in the order of their subscript numbers. Let $K$ be a piecewise function. When the position of $\vec{s}$ in sequence $A$ is $j$, $K = H_z^{-1}$, where $z$ is the integer part of the ratio of $j$ to $M\frac{N!}{N^N}$. It is evident that $K$ satisfy $\vec{t} = K(\vec{s})$, where $\vec{t} \in C$ and $\vec{s} \in A$. Because $C_1 = C_2 = \cdots = C_z = \cdots = C_{\frac{N^N}{N!}}$, $K$ is non-invertible.

Therefore, if $\vec{s}$ is selected as the independent variable of the function, we have
$$F = \psi(\vec{s}), \vec{t} = K(\vec{s}), \vec{s} \in A, \vec{t} \in C. \tag{3}$$
Therefore, if $\vec{t}$ is selected as the independent variable of the function, we have
$$\vec{s} = H_z(\vec{t}), F = \psi\left(H_z(\vec{t})\right), \vec{s} \in A_z, \vec{t} \in C_z. \tag{4}$$

The aforementioned discussion in this paper should be applicable to which of the following physical situation: the independent variables of a physical state are $s_1, s_2, s_3, \cdots, s_N$; $s_1, s_2, s_3, \cdots, s_N$ are difficult or impossible to measure, whereas $t_1, t_2, t_3, \cdots, t_N$ are easy to measure; therefore, we need to get the values of $t_1, t_2, t_3, \cdots, t_N$ by measuring first, and then we need to use the values of $t_1, t_2, t_3, \cdots, t_N$ to calculate the values of $s_1, s_2, s_3, \cdots, s_N$. Therefore, for this physical state, $s_1, s_2, s_3, \cdots, s_N$ are the independent variables, $t_1, t_2, t_3, \cdots, t_N$ are the measurable variables, and $F$ is a physical quantity. It can thus be concluded that an Equation measured in an experiment should be Equation (4), not Equation (3). Therefore, we call the state described by Equation (4) the measurable state in this paper. Because the independent variables of this physical state are $s_1, s_2, s_3, \cdots, s_N$, it is evident that $s_1, s_2, s_3, \cdots, s_N$ can take on all possible values under a complete physical state. According to Step 6, $s_1, s_2, s_3, \cdots, s_N$ can assume any of the possible values, so it can thus be concluded that the physical state described by Equation (3) is a complete physical state. According to Step 6, $s_1, s_2, s_3, \cdots, s_N$ can assume any of the possible values freely, so the physical state described by Equation (3) is called the free state in this paper. After careful study, it can be found that the domain of a free state (Equation (3) is Sequence $A$, whereas the domain of a measurable state (Equation (4)) is merely a subset of $A$ (namely $A_z$). Therefore, the measurable state is part of free state. According to Step 2, Sequence $A$ is composed of $A_1$, $A_2$, $\cdots$, $A_{\frac{N^N}{N!}}$. Thus, once this physical state has been determined, the number of free states is only one, but the number of measurable states is $\frac{N^N}{N!}$. Let $\Gamma$ be the number of measurable states in a free state, then $\Gamma = \frac{N^N}{N!}$. When $N$ is large enough, we have

$$\Gamma = e^N. \tag{5}$$

It can be seen that each of these $e^N$ measurable states is a different part of the same free state. The distinction between these $e^N$ measurable states lies in the values of their $s_1, s_2, s_3, \cdots, s_N$. According to Step 6, all possible values of $\vec{s}$ have an equal probability of occurring. Thus, the measurable states that give the same free state have the same probability. That is, the probability of a measurable state under the same free state is equal to $e^{-N}$. Let $\vartheta$ represent the probability of a measurable states occurring under the same free state. Then, we have

$$\vartheta = e^{-N}. \tag{6}$$

Measurable states and free states can be constructed in another method. Let measurable states and free states satisfy the following conditions:

**Condition 1** *For a free state, the physical quantities are functions of independent variables. That is, for this free state, $\vec{s}$ are independent variables and $\vec{t}$ are dependent variables. Then, there exists a function $K$ such that $\vec{t} = K(\vec{s})$ in this free state.*

**Condition 2** *For every measurable state in this free state, the physical quantities are functions of measurable variables. That is, for every measurable state in this free state, $\vec{t}$ are independent variables and $\vec{s}$ are dependent variables. In zth measurable state, there exists a function $H_z$ such that $\vec{s} = H_z(\vec{t})$. According to Condition 1, $\vec{t} = K(\vec{s})$ in this free state. Therefore, $H_z$ is an invertible function. Thus, in $z$th measurable state, there exists an invertible function $H_z$ such that $\vec{s} = H_z(\vec{t})$ and $\vec{t} = H_z^{-1}(\vec{s})$. It is therefore possible to construct sequence $A$ and sequence $B_z$, with sequence $A$ corresponding to the sequence shown in step 1 and sequence $B_z$ to that shown in step 3. It is the same as step 1 and step 3. A comparison of $\vec{t} = H_z^{-1}(\vec{s})$ and $A \Leftrightarrow B_z$ shows that if $A \Leftrightarrow B_z$ is applicable to a measurable state, then only the valid part of $A \Leftrightarrow B_z$ (namely $\vec{t} = H_z^{-1}(\vec{s})$) is applicable to that measurable state. It indicates that an element can be observed only when it is valid. It is the same as step 4.*

**Condition 3** *Every measurable state should be complete. It is evident that the domain of a complete measurable state should be the set of all possible measurable variables ($\vec{t}$). Then, the measurable states who give the same free state have the same domain. It is evident that condition 3 is equivalent to Step 5.*

**Condition 4** *In every measurable state, the independent variable ($\vec{t}$) is continuous. Because $\vec{s} = H_z(\vec{t})$, dependent variable ($\vec{s}$) is continuous in every measurable state. It is the same as step 8.*

**Condition 5** *For every measurable state, the relationship between independent variables ($\vec{t}$) and dependent variables ($\vec{s}$) does not change over time. It is evident that condition 5 is equivalent to Step 7.*

**Condition 6** *All the measurable states who give the same free state are distinguishable from each other.*

*Let Measurable state $A$ and Measurable state $B$ are any two Measurable states in the same free state. H function of Measurable state $A$ is $H_a$, and H function of Measurable state $B$ is $H_b$. Suppose that there exists $\vec{s}_0$ and $\vec{t}_0$ such that $\vec{s}_0 =$*

$H_a(\vec{t}_0) = H_b(\vec{t}_0)$. According to Condition 4, $\vec{s}$ is continuous in a measurable state. It therefore follows that, if $\vec{s}_0$ belongs to a measurable state, then $\vec{s}_0 + \Delta\vec{s}$ must belong to this measurable state, where $\Delta\vec{s}$ is an infinitesimal number. Therefore, there exists $\vec{t}_1$ and $\vec{t}_2$ such that $\vec{s}_0 + \Delta\vec{s} = H_a(\vec{t}_1)$ and $\vec{s}_0 + \Delta\vec{s} = H_b(\vec{t}_2)$. According to Condition 1, $\vec{t} = K(\vec{s})$, so $K$ maps every $\vec{s}$ to a unique element $\vec{t}$ in a free state. Thus, $\vec{t}_1 = \vec{t}_2 = K(\vec{s}_0 + \Delta\vec{s})$. In conclusion, we have $\vec{s}_0 + \Delta\vec{s} = H_a(\vec{t}_1) = H_b(\vec{t}_2)$.

Then, by mathematical induction, we have $\vec{s}_0 + \lambda_9\Delta\vec{s} = H_a(\vec{t}) = H_b(\vec{t})$, where $\lambda_9\$$ is any integer. Therefore, $H_a = H_b$, but it does not fit with Condition 6. So, the statement "there exists $\vec{s}_0$ and $\vec{t}_0$ such that $\vec{s}_0 = H_a(\vec{t}_0) = H_b(\vec{t}_0)$" is false.

According to Condition 1, $\vec{t} = K(\vec{s})$, so $K$ maps every $\vec{s}$ to a unique element $\vec{t}$ in a free state. So, the statement "there exists $\vec{s}_0$, $\vec{t}_0$ and $\vec{t}_3$ such that $\vec{s}_0 = H_a(\vec{t}_0) = H_b(\vec{t}_3)$, where $\vec{t}_0 \neq \vec{t}_3$" is false.

Combining these two statements shows that $\vec{s}$ cannot exist in two different Measurable states at the same time for every $\vec{s}$. In other words, for every $\vec{s}$, if $\vec{s}$ belongs to a measurable state $A$, then $\vec{s}$ does not belongs to another measurable state who give the same free state. It is evident that Condition 6 is equivalent to Step 2.

**Condition 7** In a free state, all possible values of $\vec{s}$ have an equal probability of occurring. It is the same as step 6.

**Condition 8** All measurable states obey physics laws.

The method of constructing measurable state and free state in accordance with Steps 1–7 is called method A. The method of constructing measurable state and free state in accordance with Conditions 1-8 is called method B. Conditions 1 and 2 are equivalent to step 1, step 3 and step 4. Condition 3 is equivalent to Step 5. Condition 5 is equivalent to Step 7. Condition 4 the same as step 8. Condition 6 is equivalent to Step 2. Condition 7 is the same as step 6. Therefore, method A and method B are equivalent.

It is evident that if a measurable state is a real and stable physical state, then this measurable state naturally satisfies Conditions 2, 3, 4, 5, 6 and 8. It is also evident that if a measurable state satisfies Conditions 2, 3, 4, 5, 6 and 8, it will behave like a real and stable physical state. Because method A and method B are equivalent, every measurable state satisfies Conditions 2, 3, 4, 5 and 6. Because the free state is a real physical state, the equation of a free state, $F = \psi(\vec{s})$, naturally obey the physics laws; since there is no relationship between $\vec{s}$ and $\vec{t}$, it is not possible to establish a universal relationship between $\vec{s}$ and $\vec{t}$; it is evident that the physics laws are universal; Therefore, there is no physics law between $\vec{s}$ and $\vec{t}$; thus, it is evident that any $\vec{s} = H_z(\vec{t})$ obeys the physics laws; in summary, the equation of every measurable state, $F = \psi\big(H_z(\vec{t})\big)$, obeys the physics laws, that is, every measurable state satisfies

Condition 8. In summary, every measurable state satisfies Conditions 2, 3, 4, 5, 6 and 8. Therefore, every measurable state behaves like a real and stable physical state. This is why we chose Steps 1-7.

**Example 1** *Here is an example, namely Example 1. Function $K$ is given by the equations*

$$t_i = cot(\pi s_i), i = 1,2,3,\cdots, N. \tag{7}$$

*Therefore, function $H_z$ is given by the equation*

$$s_i = \frac{cot^{-1} t_i}{\pi} + \varrho_i; i = 1,2,3,\cdots, N; t_i \in (-\infty, +\infty). \tag{8}$$

*where $\varrho_i$ is an integer and $\pi$ is the value of Pi. Then, each $(\varrho_1, \varrho_2, \varrho_3, \cdots, \varrho_N)$ represents a measurable state.*

A more thorough analysis yields the following conclusions:

1. Because the independent variables of a free state are $s_1, s_2, s_3, \cdots, s_N$, the degree of freedom of a free state is N. $t_1, t_2, t_3, \cdots, t_N$ are N mutually independent variables. Therefore, for this free state, if $t_1, t_2, t_3, \cdots, t_N$ are selected as the variables, then there is no longer an additional independent variable.

2. It is evident that a free state can only occupy one value of $(s_1, s_2, s_3, \cdots, s_N)$ at any given moment. Then according to Equation (3), this free state can only occupy one value of $(t_1, t_2, t_3, \cdots, t_N)$ at any given moment. Consequently, a single measurement of $(t_1, t_2, t_3, \cdots, t_N)$ at a given moment will yield a single result.

3. According to the previous analysis: a free state is a complete physical state and a measurable state is part of this free state; the measurable states that give the same free state have the same probability. In summary, a free state is in all possible measurable states at the same time, until it is measured. If the collision does not disturb free state, we can conclude the following: before measurement, a free state is in all possible measurable states and we are not sure which measurable state this free state is in at this time; at measurement, this free state will be observed on one of possible measurable states: after measurement, it is still in all possible measurable states and we are still not sure which measurable state this free state is in at this time. Because every measurable state behaves like a real and stable physical state, it is easy to reach some erroneous conclusions. For instance, when a physical state (free state) is not observed, it is in a superposition of some real states (measurable states) and only when the physical state is measured or observed, it then falls to one of the real states that form the superposition.

4. $s_1, s_2, s_3, \cdots, s_N$ are the independent variables of a free state. $t_1, t_2, t_3, \cdots, t_N$ is a set of measurable variables of this free state and $\chi_1, \chi_2, \chi_3, \cdots, \chi_N$ is another set of measurable variables of this free state. Then, we have

$$F = \psi(\vec{s}), \vec{\chi} = K'(\vec{s}), \vec{s} \in A, \tag{9}$$

and

$$\vec{s} = H'_z(\vec{\chi}), F = \psi(H'_z(\vec{\chi})), \vec{s} \in A_z, \tag{10}$$

where $K'$ and $H'_z$ are two function.

Once $(t_1, t_2, t_3, \cdots, t_N)$ has been determined, according to Equations (4) and (5), $(s_1, s_2, s_3, \cdots, s_N)$ has $e^N$ possible values; substituting this into Equation (9) yields that $(\chi_1, \chi_2, \chi_3, \cdots, \chi_N)$ may have more than one possible value at this case. Once

$(\chi_1, \chi_2, \chi_3, \cdots, \chi_N)$ has been determined, according to Equations (5) and (10), $(s_1, s_2, s_3, \cdots, s_N)$ has $e^N$ possible values; Substituting this into Equation (3) yields that $(t_1, t_2, t_3, \cdots, t_N)$ may have more than one possible value at this case.

Therefore, there exist two sets of measurable variables in a free state, denoted by $t_1, t_2, t_3, \cdots, t_N$ and $\chi_1, \chi_2, \chi_3, \cdots, \chi_N$. $t_1, t_2, t_3, \cdots, t_N$ and $\chi_1, \chi_2, \chi_3, \cdots, \chi_N$ satisfy: once $(t_1, t_2, t_3, \cdots, t_N)$ has been determined, $(\chi_1, \chi_2, \chi_3, \cdots, \chi_N)$ has more than one possible value; once $(\chi_1, \chi_2, \chi_3, \cdots, \chi_N)$ has been determined, $(t_1, t_2, t_3, \cdots, t_N)$ has more than one possible value. Therefore, $(t_1, t_2, t_3, \cdots, t_N)$ and $(\chi_1, \chi_2, \chi_3, \cdots, \chi_N)$ cannot both be measured exactly, at the same time. In other words, there exist two sets of measurable variables in a free state such as it is impossible to measure exactly these two sets of measurable variables at the same time.

**Example 2** *Here is an example, namely Example 2. The relationship between $t_i$, $\chi_i$ and $s_i$, for measurable state $z$ ($z = 1,2,3,\cdots, e^N$), is given by the equations*

$$t_i = \tan s_i, s_i = \tan^{-1}(t_i) + z\pi, \chi_i = s_i \tan s_i, \qquad (11)$$

*where $s_i \neq 0$ and $i = 1,2,3,\cdots, N$. For any two different $s_i$, namely $s_{i1}$ and $s_{i2}$, where $s_{i1} \neq s_{i2}$. Because $s_i \neq 0$, $s_{i1} \neq s_{i2} \neq 0$. Therefore, if $\tan s_{i1} = \tan s_{i2}$, then $s_{i1}\tan s_{i1} \neq s_{i2}\tan s_{i2}$; if $s_{i1}\tan s_{i1} = s_{i2}\tan s_{i2}$, then $\tan s_{i1} \neq \tan s_{i2}$. Thus, once $t_i$ ($\tan s_i$) has been determined, $\chi_i$ ($s_i \tan s_i$) has more than one possible value; once $\chi_i$ ($s_i \tan s_i$) has been determined, $t_i$ ($\tan s_i$) has more than one possible value. In other words, it is impossible to measure exactly $(t_1, t_2, t_3, \cdots, t_N)$ and $(\chi_1, \chi_2, \chi_3, \cdots, \chi_N)$ at the same time.*

## 3 A NEW TRANSITION

It is evident that the most common measurable variables are the spatial positions. Let $U$ be a free state. Free state $U$ satisfies:

**U₁** *The measurable variables of free state $U$ are the spatial positions.*

**U₂** *The systems discussed in this section are limited to those consisting of $X$ independent mass points whose dimensions and deformations are negligible. It is evident that, a $3X$-dimensional coordinate system can be used to represent the spatial positions of $X$ mass points. Therefore, the measurable variables of free state $U$ for a system is a set of $3X$ numbers that represent the spatial positions of $X$ mass points. According to Equation (5), for this system, the free state $U$ has $e^{3X}$ measurable states. Let $\Gamma(U)$ be the number of measurable states in free state $U$ for a system. Thus, $\Gamma(U) = e^{3X}$.*

**U₃** *Every measurable state in free state $U$ for a system can be represented as $(b_1, b_2, b_3, \cdots, b_N)$, where $b_i$ is the characteristic value of the $i$th mass point. It is similar to Example 1, in which a measurable state can be represented as $(\varrho_1, \varrho_2, \varrho_3, \cdots, \varrho_N)$. Meanwhile, the free state $U$, for a system, allows each mass point to assume $\varpi$ different characteristic value. Therefore, there are $\varpi^X$ possible values for $(b_1, b_2, b_3, \cdots, b_N)$. Thus, for a system, the free state $U$ has $\varpi^X$ measurable states. That is $\Gamma(U) = \varpi^X$.*

Combining $\Gamma(U) = e^{3X}$ and $\Gamma(U) = \varpi^X$, we have $e^{3X} = \varpi^X$. Thus, we have

$$\varpi = e^3. \tag{12}$$

$e^3$ is an irrational number. It suggests that $e^3$ may be a statistical average.

According to U3, the free state $U$, for a system, allows each mass point to assume $\varpi$ different characteristic value. In other words, $b_i$ can assume $\varpi$ different values. Let the possible values of $b_i$ be $b_{i1}, b_{i2}, b_{i3}, \cdots, b_{i\varpi}$. It is evident that once the values of $b_{i1}, b_{i2}, b_{i3}, \cdots, b_{i\varpi}$ have been determined, all possible values of $b_i$ are determined. It follows that once the values of $b_{11}, b_{12}, \cdots, b_{ij}, \cdots, b_{X\varpi}$ have been determined, all possible values of $(b_1, b_2, b_3, \cdots, b_N)$ are determined, where $i \leq X$ and $j \leq \varpi$. Every measurable state in free state $U$ for a system can be represented as $(b_1, b_2, b_3, \cdots, b_N)$. Thus, once the values of $(b_{11}, b_{12}, \cdots, b_{ij}, \cdots, b_{X\varpi})$ have been determined, all measurable state in free state $U$ for a system are determined such that free state $U$ for a system is determined. In consequence, the variables of free state $U$ are $b_{11}, b_{12}, \cdots, b_{ij}, \cdots, b_{X\varpi}$.

Let $\theta$ motion represent the motion of a system from one $U$ free state to another $U$ free state. Because the variables of free state $U$ are $b_{11}, b_{12}, \cdots, b_{ij}, \cdots, b_{X\varpi}$, $\theta$ motion is the motion of a system from a $(b_{11}, b_{12}, \cdots, b_{ij}, \cdots, b_{X\varpi})$ to another. Therefore, $\theta$ motion can represent as $\Delta b_{11}, \Delta b_{12}, \cdots, \Delta b_{ij}, \cdots, \Delta b_{X\varpi}$, where $\Delta b_{ij}$ represents the difference in $b_{ij}$ between the initial and final states of $\theta$ motion. Let $V$ be the free state that describe the $\theta$ motion of a system. Thus, the variables of free state $V$ for a system are $\Delta b_{11}, \Delta b_{12}, \cdots, \Delta b_{ij}, \cdots, \Delta b_{X\varpi}$. Therefore, the variables of free state $V$ for a system is a set of $X\varpi$ ($Xe^3$) numbers. According to Equation (5), for a system, the free state $V$ has $e^{Xe^3}$ measurable states. Let $\Gamma(V)$ be the number of measurable states in free state $V$ for a system. That is,

$$\Gamma(V) = e^{Xe^3}. \tag{13}$$

The transition of an electron from one stationary state to another is typically accompanied by a change in its energy. Inspired by this, it is assumed that the transition of a mass point from one state $V$ to another state $V$ may be accompanied by a change its spatial position. In other words, it is assumed that $\theta$ motion of a mass point may be accompanied by a spatial motion of this mass point. The motion that accompanies $\theta$ motion is called $\xi$ motion. Accordingly, the observation of $\theta$ motion of a mass point will yield the following three results: the $\theta$ motion is not accompanied by $\xi$ motion for this mass point; the $\theta$ motion is accompanied by $\xi$ motion for this mass point, but this mass point is not in $\xi$ motion at the moment of observation; the $\theta$ motion is accompanied by $\xi$ motion for this mass point, and this mass point is in $\xi$ motion at the moment of observation. Therefore, the mass points can divide into two types: the first type is not accompanied by $\xi$ motion and the second type is accompanied by $\xi$ motion. The second type can also divide into two types: the third type is not in $\xi$ motion and the fourth type is in $\xi$ motion.

Let $\xi$ motion satisfy:

**D1** *According to the analysis in section 2, a free state is a complete physical state*

*and a measurable state is part of this free state. The thermal behavior of a physical state is described in terms of thermodynamic variables and the thermodynamic variables have unique values for a physical state. Therefore, the thermodynamic variables have unique values for free state V.*

**D₂** *A mass point in the third type is in the same phase as this mass point in fourth type. That is, under the same conditions, the values of every thermodynamic variable of a substance in the third type is equal to this of this substance in the fourth type. By combining D₁ and D₂, it can be concluded that the ratio of the third type of mass points to the fourth type of mass points in free state V can assume any value.*

**D₃** *A mass point in the first type is not in the same phase as this mass point in second type. That is, under the same conditions, there is at least one thermodynamic variable such that the value of this thermodynamic variable for a substance in the first type is different from those of this substance in the second type. By combining D₁ and D₃, it can be concluded that the ratio of the first type of mass points to the second type of mass points in free state V has a unique value.*

Accordingly, free state $V$ is selected as "$X - L$ of $X$ mass points are not accompanied by $\xi$ motion, the remaining $L$ mass points are accompanied by $\xi$ motion". Event $\Xi$ under the free state $V$ is as follows: $X - L$ of $X$ mass points are not accompanied by $\xi$ motion, the remaining $L$ mass points are accompanied by $\xi$ motion; $H$ of the remaining $L$ mass points are in $\xi$ motion, and $L - H$ of the remaining $L$ mass points are not in $\xi$ motion, where $H \leq L$. The term ``molecule" is used to represent the smallest independent unit of $\xi$ motion. Let $W$ be the number of mass points in a molecule and $p$ be the probability that a molecule is in $\xi$ motion. Therefore, Event $\Xi$ is equivalent to as follows: $\frac{X-L}{W}$ of $\frac{X}{W}$ molecules are not accompanied by $\xi$ motion; $\frac{H}{W}$ of the remaining $\frac{L}{W}$ molecules are in $\xi$ motion, and $\frac{L-H}{W}$ of the remaining $\frac{L}{W}$ molecules are not in $\xi$ motion. Then, the probability of Event $\Xi$ occurring under the free state $V$ is $\binom{\frac{L}{W}}{\frac{H}{W}} p^{\frac{H}{W}} (1-p)^{\frac{L-H}{W}}$, where $\binom{\frac{L}{W}}{\frac{H}{W}}$ is the binomial coefficient.. It is evident that a necessary condition for Event $\Xi$ occurring is that this event corresponds to at least one measurable state. Therefore, event $\Xi$ can occur only if $\binom{\frac{L}{W}}{\frac{H}{W}} p^{\frac{H}{W}} (1-p)^{\frac{L-H}{W}} \geq \frac{1}{\Gamma(V)}$. Substituting this into Equation (14) yields that

$$\binom{\frac{L}{W}}{\frac{H}{W}} p^{\frac{H}{W}} (1-p)^{\frac{L-H}{W}} \geq e^{-X e^3}. \tag{14}$$

Because $H \leq L$, $\binom{\frac{L}{W}}{\frac{H}{W}} p^{\frac{H}{W}} (1-p)^{\frac{L-H}{W}} > 0$ for all $H$. That is, the probability of

every event Ξ occurring under the free state $V$ is not zero. Therefore, every event Ξ under the free state $V$ is possible. Accordingly, if the free state $V$ exists, then every event Ξ under the free state $V$ is possible. In other words, a necessary condition for the existence of the free state $V$ is that equation (14) holds for every event Ξ in free state $V$. It is evident that $\begin{pmatrix} \frac{L}{W} \\ \frac{H}{W} \end{pmatrix} p^{\frac{H}{W}}(1-p)^{\frac{L-H}{W}} \geq p^{\frac{L}{W}}$. In summary, free state $V$ can exist only if $p^{\frac{L}{W}} \geq e^{-Xe^3}$. Therefore, we have

$$1 - \frac{L}{X} \geq 1 + \frac{We^3}{\ln p}. \tag{15}$$

Let $G$ be $1 - \frac{L}{X}$. In other words, $G$ is equal to the proportion of molecules that are not accompanied by ξ motion. Then, according to Equation (15), we have

$$G \geq 1 + \frac{We^3}{\ln p}. \tag{16}$$

Let $G_{min}$ be the minimum value of $G$. It is evident that $0 \leq G \leq 1$. Substituting these into Equation (16) yields that when $p \geq e^{-We^3}$,

$$G_{min} = 0; \tag{17}$$

and when $p < e^{-We^3}$

$$G_{min} = 1 + \frac{We^3}{\ln p}. \tag{18}$$

According to Equations (17) and (18), $G_{min} = 0$ when $p \geq e^{-We^3}$ and $G_{min} > 0$ when $p < e^{-We^3}$. Therefore, there is a transition here, and at the transition point

$$p = e^{-We^3}. \tag{19}$$

In light of the fact that the most distinctive feature of this transition is $e^3$, this paper will henceforth refer to it as the $e^3$ transition.

The molecule is the smallest independent unit of ξ motion. If the molecule does not change shape in ξ motion, then it can be regarded as a particle or rigid body in ξ motion. It is evident that, a 3-dimensional coordinate system can be used to represent the spatial positions of a particle and a 6-dimensional coordinate system can be used to represent the spatial positions of a rigid body. According to U₂, a $3X$-dimensional coordinate system can be used to represent the spatial positions of $X$ mass points. Therefore, there is one mass point in a particle, while there are two mass points in a rigid body. $W$ is the number of mass points in a molecule. In conclusion, when the molecule can be regarded as a particle in ξ motion, $W = 1$. Substituting these into Equation (19) yields that at the transition point,

$$p = e^{-e^3}. \tag{20}$$

If the molecule can be regarded as a particle in a $\xi$ motion, this $\xi$ motion will be called $\beta$ motion. If the molecule can be regarded as a rigid body in a $\xi$ motion, this $\xi$ motion will be called $\alpha$ motion. Similarly, for a molecule in $\alpha$ motion, $W = 2$ and at the transition point,

$$p = e^{-2e^3}. \tag{21}$$

## 4  $e^3$ TRANSITION OF LIQUIDS

Assume that the $e^3$ transition will occur in liquid. According to Equation (20), if the liquid molecule is regarded as a particle in $\xi$ motion (that is, if the liquid molecule is in $\beta$ motion), $p_1 = e^{-e^3}$ at the transition point, where $p_1$ is the probability that a molecule is in $\beta$ motion. According to Equation (21), if the liquid molecule is regarded as a rigid body in $\xi$ motion (that is, if the liquid molecule is in $\alpha$ motion), $p_2 = e^{-2e^3}$ at the transition point, where $p_2$ is the probability that a molecule is in $\alpha$ motion. It can be seen that the liquid will undergo two transitions. One transition points is observed at $-\ln(p_1) = e^3$, and another transition is observed at $-\ln(p_2) = 2e^3$. Therefore, $e^3$ and $2e^3$ are the characteristic values for these two transitions. Next, we will check whether these two characteristic values are valid or not.

### 4.1  $2e^3$

In accordance with the findings of Weihua Wang and his partners, the RFV (reduced free volume) of MGs (metallic glasses) sharply increases (becomes diverge) when it reaches a critical value of about 2.4% for the onset of yielding [12,31]. It is evident that $\frac{1}{2e^3} \approx 0.025$. In summary, the critical value of RFV for the onset of yielding is very close to $\frac{1}{2e^3}$.

The Vogel–Fulcher–Tammann equation (VFT equation) is used to describe the temperature dependence of the viscosity ($\eta$) or of the relaxation time in various types of supercooled liquids including metallic glass forming materials. The VFT expression of the viscosity is given by $\eta = \lambda_3 \exp\left(\frac{\lambda_4 T_0}{T-T_0}\right)$, where $\lambda_3$ and $\lambda_4$ are constants, $T_0$ is VFT temperature and $T$ is the temperature [32]. The VFT equation is mathematically equivalent to Vogel–Fulcher equation ($\eta = \lambda_5 \exp\left(\frac{\lambda_6}{T-T_0}\right)$, VF equation) and Williams-Landel-Ferry equation ($\eta= \exp\left(\frac{\lambda_1(T-T_g)}{\lambda_2+T-T_g}\right)$, WLF equation) [32], where $\lambda_1$, $\lambda_2$, $\lambda_5$, $\lambda_6$ and $\lambda_7$ are constants. Here, the fractional free volume, $f$, is defined as

$$f = \frac{T-T_0}{\lambda_4 T_0} = \frac{T-T_0}{\lambda_6} = \frac{1}{\lambda_1} + \frac{T-T_g}{\lambda_1 \lambda_2}. \tag{22}$$

Table 1 lists the fractional free volume at the glass transition temperature for a number of materials, where $f_g$ is the fractional free volume at the glass transition temperature. These materials include alloys, small molecules, polymers, etc. It can be observed that $f_g$ of these materials is about 0.025.

In summary, $f_g$ appears to be about 0.025 for materials. Because $\frac{1}{2e^3} \approx 0.025$, $f_g$ appears to be about $\frac{1}{2e^3}$ for materials.

**TABLE 1** The $f_g$ of various materials

| Name | $f_g$ |
|---|---|
| **Sodium-silicate glasses $Na_2O - SiO_2$** | |
| 15 $Na_2O$–85 $SiO_2$[33] | 0.028 |
| 20 $Na_2O$–80 $SiO_2$[33] | 0.028 |
| 25 $Na_2O$–75 $SiO_2$[33] | 0.028 |
| 30 $Na_2O$–70 $SiO_2$[33] | 0.028 |
| 33 $Na_2O$–67 $SiO_2$[33] | 0..028 |
| 35 $Na_2O$–65 $SiO_2$[33] | 0.028 |
| **Poly-alkali silicate glass** | |
| 69.04$SiO_2$ 30.96$Na_2O$ [33] | 0.022 |
| 79.29$SiO_2$ 12.97$Na_2O$ 7.75$Li_2O$ [33] | 0.022 |
| 43.22$SiO_2$ 9.55$Na_2O$ 47.23$CsO$ [33] | 0.032 |
| 71.59$SiO_2$ 24.4$Na_2O$ 4.01$Li_2O$ [33] | 0.028 |
| **Amorphous polymers** | |
| Polyisobutylene [33] | 0.026 |
| Polyvinyl acetate [33] | 0.028 |
| Polyvinyl chloroacetate[33] | 0.025 |
| Polymethyl acrylate[33] | 0.024 |
| Polyurethane[33] | 0.028 |
| Natural rubber[33] | 0.026 |
| Methacrylate polymers ethyl[33] | 0.025 |
| Methacrylate polymers n-butyl[33] | 0.026 |
| Methacrylate polymers n-octyl[33] | 0.027 |
| $BMIMBF_4$[34] | 0.028 |
| TMPTos[34] | 0.031 |
| PBI[35] | 0.029 |
| PBT[35] | 0.030 |
| **Metallic glass** | |
| $Pd_{40}Ni_{40}P_{20}$[33] | 0.026 |
| $Pt_{60}Ni_{15}P_{25}$[33] | 0.027 |
| $Pd_{77.5}Cu_6Si_{16.5}$[33] | 0.026 |
| $Fe_{80}P_{13}C_7$[33] | 0.026 |
| Selenium [33] | 0.031 |
| $Pd_{40}Cu_{30}Ni_{10}P_{20}$[36] | 0.027 |
| Zr$_{55}$Cu$_{30}$Al$_{10}$Ni$_5$[36] | 0.027 |
| Zr$_{46.75}$Ti$_{8.25}$Cu$_{7.5}$Ni$_{10}$Be$_{22.5}$[36] | 0.032 |
| Mg$_{65}$Cu$_{25}$Y$_{10}$[36] | 0.026 |
| Pd$_{40}$Ni$_{40}$P$_{20}$[36] | 0.024 |
| Zr$_{55}$Cu$_{30}$Al$_{10}$Ni$_5$[36] | 0.027 |
| Pd$_{40}$Cu$_{30}$Ni$_{10}$P$_{20}$[36] | 0.027 |
| Fe$_{40}$Ni$_{40}$P$_{14}$B$_6$[36] | 0.028 |
| Al$_{87}$Ni$_7$Ce$_6$[36] | 0.027 |
| Vit$_4$[37] | 0.028 |
| La$_{57.5}$Al$_{17.5}$Ni$_{12.5}$Cu$_{12.5}$[37] | 0.027 |
| **Low molecular weight organic glass** | |
| Propanol[33] | 0.024 |
| Prothylene Glycol[33] | 0.023 |
| Glycerol[33] | 0.024 |
| **Strong glass** | |
| HQGeO$_2$[38] | 0.022 |

### 4.2 $e^3$

V. N. Novikov and A. P. Sokolov have collected $T_1$ (the critical temperature of mode-coupling theory) values for 26 different systems from various literature sources[39]. These systems include molecular and polymeric glass formers, covalent, hydrogen-bonded, ionic, van der Waals and disordered systems [39]. From the literature data, it can be seen that $T_1$ seems to roughly $1.2T_g$. It has been found that a considerable number of dynamic properties of glass-forming liquids undergo alterations at temperatures proximate to $T_1$[39]. Notwithstanding the existence of counterexamples, the majority of glass-forming systems display a decoupling of the $\alpha$-relaxation (main structural relaxation) and slow $\beta$-relaxation (secondary relaxation) processes at temperatures proximate to $T_1$ [39]. Furthermore, decoupling of rotational and translational diffusion has been observed at temperatures proximate to $T_1$ [39]. Also, a liquid-liquid transition will occur at a temperature of approximately $1.2T_g$ [40]. In summary, liquids will appear to undergo a transition at a temperature of roughly $1.2T_g$. Let $T_\beta$ be at the transition temperature of this transition. Then, $T_\beta$ appears to be concentrated near $1.2T_g$.

For the Johari-Goldstein $\beta$ relaxation (JG $\beta$ relaxation or slow $\beta$ relaxation) in the glassy state, we have $\tau_{JG} = \tau_\infty \exp\left(\frac{E_\beta}{RT}\right)$, where $\tau_{JG}$ is the JG $\beta$ relaxation time, $\tau_\infty$ is the prefactor and $E_\beta$ is the activation enthalpy of JG $\beta$ relaxation [19]. By analysing more than 20 nonmetallic glasses, Kudlik et al. found that the $E_\beta$ of these different glassy bodies is concentrated near $24RT_g$ [19]. Weihua wang et al. found that a similar empirical relationship between $E_\beta$ and $T_g$ in the form of $E_\beta \approx (26 \pm$

2) $RT_g$ also exists in metallic glasses [21]. For HQGeO$_2$, the activation energy $E_\beta$ is equal to $23.5RT_g$ [38]. In summary, $E_\beta$ appears to be concentrated near $24RT_g$ for materials.

In summary, $T_\beta$ and $E_\beta$ appears to be concentrated near $1.2T_g$ and $24RT_g$ for materials, respectively. Therefore, $\frac{E_\beta}{RT_\beta}$ appears to be concentrated near 20. Because $e^3 \approx 20$, $\frac{E_\beta}{RT_\beta}$ appears to be concentrated near $e^3$.

A large amount of data indicate that liquids will undergo two phase transitions. For the first transition, $\frac{E_\beta}{RT_\beta}$ appears to be concentrated near $e^3$ and for the second transition, $\frac{1}{f_g}$ or $\frac{1}{RFV}$ appears to be about $2e^3$. Although $e^3$ and $2e^3$ are only approximate values for $\frac{E_\beta}{RT_\beta}$ and $\frac{1}{f_g}$, respectively, and there are significant discrepancies between $\frac{E_\beta}{RT_\beta}$ and $e^3$ (or $\frac{1}{f_g}$ and $2e^3$) for individual substances, this finding is still interesting and significant. This finding intuitively proves that the two numbers $e^3$ and $2e^3$ play a decisive role in determining the values of $\frac{E_\beta}{RT_\beta}$ and $\frac{E_\beta}{RT_\beta}$, respectively, and it also suggests that the physical nature of $\frac{E_\beta}{RT_\beta}$ and the physical nature of $\frac{E_\beta}{RT_\beta}$ may be unified. This finding also shows that the characteristic values calculated by Equation (19) are valid. Nevertheless, the characteristic values calculated by Equation (19) is not without flaws. The primary issue with Equation (19) is that it suggests that the values of $\frac{E_\beta}{RT_\beta}$ and $\frac{1}{f_g}$ are constants, whereas in fact, the values of $\frac{E_\beta}{RT_\beta}$ and $\frac{1}{f_g}$ differ for each material. This indicates that equation (19) is relatively straightforward and that a single parameter, $p$, is insufficient for fully describing the transitions of liquids. Consequently, it is necessary to add more parameters to equation (19).

## 5    CONCLUSIONS

In this paper, concepts such as measurable state and free state are proposed, and the following conclusions are drawn:

1. The free state is a complete physical state and the measurable state is part of a free state. the number of measurable states in a free state is $e^N$. The measurable states that give the same free state have the same probability. A free state is in all possible measurable states at the same time, until it is measured.

2.There exist two sets of measurable variables in a free state such as it is impossible to measure exactly these two sets of measurable variables at the same time.

3. There may exist a special transition whose characteristic equation is $G \geq 1 + \frac{We^3}{\ln p}$. If the special transition will occur in liquids, $e^3$ and $2e^3$ are characteristic values of this special transition of liquids. A large amount of experimental data indicate that liquids will undergo two transitions. For the first transition, $\frac{E_\beta}{RT_\beta}$ appears to be concentrated near $e^3$ and for the second transition, $\frac{1}{f_g}$ or $\frac{1}{RFV}$ appears to be about $2e^3$. It shows that the two characteristic values, $e^3$ and $2e^3$, are valid.